# Gradient zigzag metamaterial beams as broadband vibration isolators for beam-like structures

**Jun Zhang** [a,b,c*], **Xuebin Zhang**[a], **Han Zhang**[d,e], **Xiaoyang Bi**[e], **Ning Hu**[e*], **Chuanzeng Zhang**[c]

[a] *College of Aerospace Engineering, Chongqing University, Chongqing, 400044, China*

[b] *Chongqing Key Laboratory of Heterogeneous Material Mechanics, Chongqing University, Chongqing, 400044, China*

[c] *Department of Civil Engineering, University of Siegen, D-57068 Siegen, Germany*

[d] *Key Laboratory of Noise and Vibration, Institute of Acoustics, Chinese Academy of Sciences, Beijing 100190, China*

[e] *Department of Mechanical Engineering, Hebei University of Technology, Tianjin, 300401, China*

**Abstract**: Phononic crystals (PCs) and metamaterials are artificially structured materials with the unprecedented property of the existence of complete stop-bands. However, the generally narrow width of the stop-bands in such materials severely limits their applicability. Here, a particular kind of gradient zigzag metamaterial beams exclusively with flat bands within a broad frequency range is proposed based upon our previously proposed uniform counterparts. The calculated band structures by the Finite Element Method (FEM) and the Spectral Element Method (SEM) show the distinctly different characteristics of the gradient zigzag metamaterial beams from the conventional PCs and metamaterials, where the band structures are totally filled with flat bands within a broad frequency range without the normal pass bands. These flat bands are caused by the local resonance of certain parts in the gradient zigzag metamaterial beams, and the elastic waves with the frequencies at these flat bands are trapped in the resonant parts and cannot propagate. As a result, our proposed gradient zigzag metamaterial beams can be used as broadband vibration isolators for beam-like structures, which is verified by both numerical simulations and experiments conducted on the 3D-printed samples. This design strategy opens a new avenue for designing and constructing novel broadband vibration isolators.

resonance, 3D-printing

# 1 Introduction

Control of acoustic and elastic waves is a lasting hot topic and mainly based upon two mechanisms: one is the Bragg scattering and the other is the local resonance. The artificially structured materials devised by means of the first mechanism is called phononic crystals [1, 2] and those designed according to the latter mechanism is usually termed metamaterials [3]. One of the most notable characteristics for both of them is the existence of complete or directional stop-bands or band-gaps, within which no acoustic or elastic waves can propagate. Such a superior property can be exploited in many applications ranging from wave filters [4, 5] to vibration isolation [6] and vibration energy harvesting [7, 8]. To date, although numerous PCs and metamaterials have been proposed, the relatively narrow bandwidth of the stop-bands still severely limits their applicability. Consequently, broadening stop-bands for both PCs and metamaterials still highly attracts the attention of researchers in various fields.

Currently, topology optimization has become the dominant technique to design PCs [9-11] with wider stop-bands. However, for metamaterials, although some [12-14] of them have been designed also using the topology optimization techniques, a more prevailing routine to enlarge the width of stop-bands is to tune or design the resonances of the employed local resonators by experiences. The proposed approaches include the active control [4, 15-20], the employment of more dissimilar resonators inside a unit-cell [21-27], and the utilization of staked arrays of resonators [8, 28-30] operating at gradually varying frequencies. It is worthy to note that, although the various active control techniques have been successfully used to increase the operating frequency ranges of metamaterials, their contribution to the widening of stop-bands for a specific control parameter is still limited, which is certainly not beneficial to the vibration isolation under a broadband excitation.

Specific to the control of elastic waves in beam-shaped structures [31-34], which is also the objective of our present study, the topology optimization based broadband

PCs have yet not been extensively reported in literature although the topology optimization has been widely used to design two-dimensional (2D) PCs [10, 35, 36] for in-plane and out-of-plane elastic waves. In contrast to conventional PCs, numerous broadband metamaterial beams have been proposed with various local resonators, including spring-mass resonators [15, 22, 24-26, 30], cantilever beam oscillators [18-21, 28], and so forth. In general, the stop-bands induced by uniform resonators are extremely narrow, hence, the overall band-gap spectrum generated by different or gradient resonators is discrete and discontinuous unless the metamaterial is constituted by a large number of resonators with adjoining working frequencies. However, this requirement is impractical in most applications. Therefore, the straightforward question is whether it is possible to design and realize a metamaterial or a PC beam with a continuous resonance frequency spectrum.

In order to attempt to answer the above question, in this work, we propose a kind of gradient zigzag metamaterial beam based upon our previously proposed uniform counterparts [37]. Compared with the uniform, the band structure of this gradient one is exclusively constituted by flat bands in a broad frequency range, and these flat bands are caused by local resonance of certain parts in the gradient metamaterial beam, although it is devised without any explicitly designed resonators as the conventional metamaterials. As a straightforward application, our proposed gradient zigzag metamaterial beams can be used as broadband wave or vibration isolators for beam-like structures.

## 2  Description of gradient zigzag metamaterial beams

In this work, upon the uniform zigzag metamaterial beam (**Fig. 1** (a)) proposed in our previous work [37], we design a gradient counterpart as presented in **Fig. 1**(b). The gradient zigzag metamaterial beam is composed of six ∏-shaped components with a gradually changing height. The graded height changes as follows

$$H_i = \left[1 + (i-1)\times \alpha\right]\times 2h, \ (i = 1, 2, 3, 4, 5, 6), \tag{1}$$

where $H_i$ represents the height of the $i^{\text{th}}$ component, $\alpha$ is a constant which determines the gradient level, and $h$ denotes the thickness of the host metamaterial beam. In this work, the geometrical parameters $h$ and $l$ are selected as $h = 2.0$ mm and $l = 4.0$ mm, unless otherwise stated, although there is no limitation to take other values. Then, the lattice constant is $a=2*(h+l)=12.0$ mm for the uniform case, and the corresponding lattice constant for the gradient case is six times $a$. In order to facilitate the fabrication of the experimental samples, the host medium is selected as a 3D-printing material (VeroPureWhite [37]) with a measured Young's modulus of $E= 3.2394$ GPa, Poisson's ratio $v=0.4185$, and mass density $\rho=1185$ kg/m$^3$ (see Supplementary material 1).

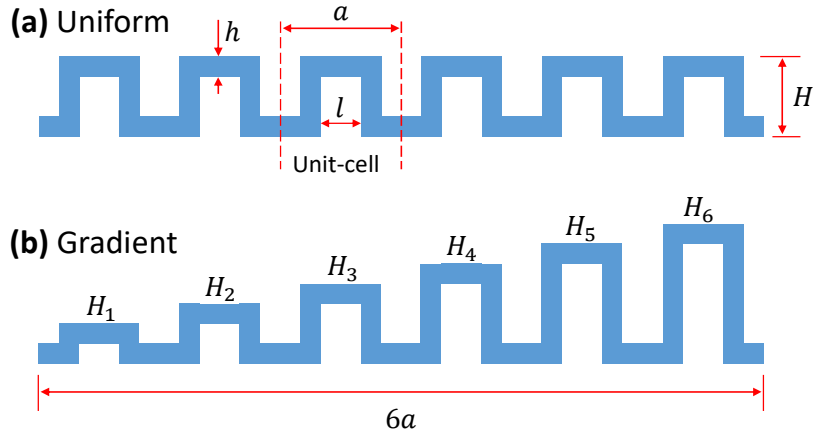

**Fig. 1**. Two-dimensional (2D) schematics of (a) the uniform and (b) the gradient zigzag metamaterial beams. For the gradient case, the height for each $\Pi$-shaped component is given on its top.

## 3 Calculation and analysis of the band structures

*3.1 Problem formulation and characteristics of the calculated band structures*

In order to characterize the behavior of the gradient zigzag metamaterial beams, their real and complex band structures are calculated using the FEM and the SEM, respectively. The numerical simulations using the FEM are conducted by the software COMSOL Multiphysics 5.4, Solid Mechanics Module. The SEM is self-implemented based on the numerical model as presented in **Fig. 2**. After the discretization of the model using Euler-Bernoulli beam theory, the discrete system of algebraic equations can be obtained as follows

$$\begin{bmatrix} S_{11} & S_{12} & S_{13} \\ S_{21} & S_{22} & S_{23} \\ S_{31} & S_{32} & S_{33} \end{bmatrix} \begin{bmatrix} \mathbf{u}_1 \\ \mathbf{u}_I \\ \mathbf{u}_{26} \end{bmatrix} = \begin{bmatrix} \mathbf{F}_1 \\ 0 \\ \mathbf{F}_{26} \end{bmatrix}, \quad (2)$$

where **S** represents the coefficient matrix whose detailed expressions can be found in Ref.[38], **u** and **F** denote the nodal displacement and force vectors, respectively. The subscripts 1 and 26 denote the values at the nodes 1 and 26, respectively, and the subscript $I$ represents the values at the interior nodes 2-25.

From Eq. (2), $\mathbf{u}_I$ can be expressed in terms of $\mathbf{u}_1$ and $\mathbf{u}_{26}$ as $\mathbf{u}_I = -S_{22}^{-1}(S_{21}\mathbf{u}_1 + S_{23}\mathbf{u}_{26})$, and after eliminating $\mathbf{u}_I$ Eq. (2) can be rewritten as

$$\begin{bmatrix} S_{11} & S_{12} & S_{13} \\ S_{31} & S_{32} & S_{33} \end{bmatrix} \begin{bmatrix} \mathbf{I}_{3*3} & 0 \\ -S_{22}^{-1}S_{21} & -S_{22}^{-1}S_{23} \\ 0 & \mathbf{I}_{3*3} \end{bmatrix} \begin{bmatrix} \mathbf{u}_1 \\ \mathbf{u}_{26} \end{bmatrix} = \begin{bmatrix} \mathbf{F}_1 \\ \mathbf{F}_{26} \end{bmatrix} \quad (3)$$

where **I** is the unit matrix. The Bloch theorem states that there exists the following relation

$$\begin{aligned} \mathbf{u}_{26} &= \mathbf{u}_1 e^{\iota ka}, \\ \mathbf{F}_{26} e^{-\iota ka} &+ \mathbf{F}_1 = 0, \end{aligned} \quad (4)$$

Here, $\iota = \sqrt{-1}$ is the imaginary unit. Then, substituting Eq. (4) into Eq. (3) gives

$$\begin{bmatrix} \mathbf{I}_{3*3} & \frac{1}{\lambda}\mathbf{I}_{3*3} \end{bmatrix} \begin{bmatrix} S_{11} & S_{12} & S_{13} \\ S_{31} & S_{32} & S_{33} \end{bmatrix} \begin{bmatrix} \mathbf{I}_{3*3} & 0 \\ -S_{22}^{-1}S_{21} & -S_{22}^{-1}S_{23} \\ 0 & \mathbf{I}_{3*3} \end{bmatrix} \begin{bmatrix} \mathbf{I}_{3*3} \\ \lambda \mathbf{I}_{3*3} \end{bmatrix} \mathbf{u}_1 = 0, \quad (5)$$

where $\lambda = e^{\iota ka}$. Equation (5) can be expressed in terms of $\lambda$ as

$$(\mathbf{D}_2 \lambda^2 + \mathbf{D}_1 \lambda + \mathbf{D}_0) \mathbf{u}_1 = 0 \quad (6)$$

with the coefficient matrices $\mathbf{D}_j$ ($j = 0, 1, 2$), whose detailed expressions are given in the Supplementary material 2. Then the eigenvalues $\lambda$ for each given $\omega$ can be obtained by using the polynomial eigenvalue solver *polyeig* in MATLAB.

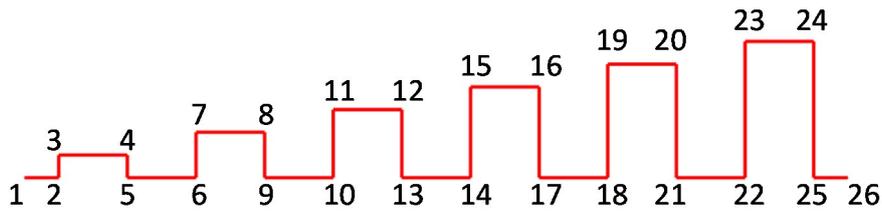

**Fig. 2**. The schematic of numerical models for the SEM to calculate the complex band

structures of the zigzag metamaterial beams. The numbers 1-26 denote the numbers of the 26 nodes.

For comparison, the corresponding results for the uniform cases of different heights are also calculated based upon the super-cell consisting of 6 unit-cells. In the FEM, 2D plane-stress solid elements are utilized to discretize the super-cell. Here, the FEM results can be considered as the benchmark solutions. A schematic of the numerical models for the uniform and the gradient cases in the SEM modeling is depicted in **Fig. 3**, in which each straight line represents an Euler-Bernoulli beam and a spectral element.

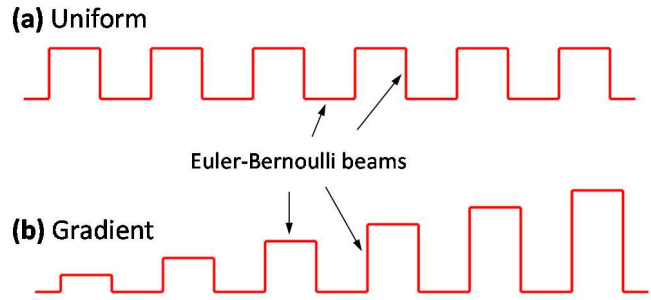

**Fig. 3**. Numerical models used to calculate complex band structures by the SEM for (a) the uniform case and (b) the gradient case.

**Fig. 4** shows the calculated real and imaginary parts of the band structures for the uniform case and the gradient case with a value $\alpha = 0.5$. For this considered specific gradient case, the heights for the six components are $H_1 = 2h$, $H_2 = 3h$, $H_3 = 4h$, $H_4 = 5h$, $H_5 = 6h$, and $H_6 = 7h$. As expected, the SEM results are in better agreements with those of the FEM at low frequencies than at high frequencies, since the employed Euler-Bernoulli beam theory in the SEM is more reasonable at lower frequency ranges. From **Fig. 4**, we can find that there exists an essential difference between the band structures of the gradient case and those of the uniform one. That is, the real part of the band structures of the gradient case is exclusively constituted by a set of flat bands (horizontal dispersion curves) as the frequency is above about 2.0 kHz in this case, without any other normal pass bands as in the uniform case. Such a difference is also reflected by the existence of a sequence of adjoining semi-arcs in the corresponding imaginary parts of the band structures in the bottom right sub-figure of **Fig. 4**. In the sub-figures of **Fig. 4**(b), the frequency range in which the imaginary part of wavenumbers is not equal to zero represents a stop band, otherwise it denotes a pass band. In addition, the elastic waves corresponding to the flat bands have a group velocity of $c_g = d\omega/dk \approx 0$. **Fig. 5** plots the calculated eigen-modes corresponding to the four representative flat bands denoted by the black arrows A, B, C and D in the rightmost sub-figure of **Fig. 4**(a). The results confirm that these flat bands are caused by the local resonance of certain parts in the metamaterial beam. Although flat bands have been observed in some previously proposed PCs by other researchers [39-41], however, to the best of our knowledge, metamaterials and/or PCs exclusively exhibiting multiple flat bands within a relatively wide frequency range as our proposed gradient zigzag metamaterial beams have not yet been reported in literature so far.

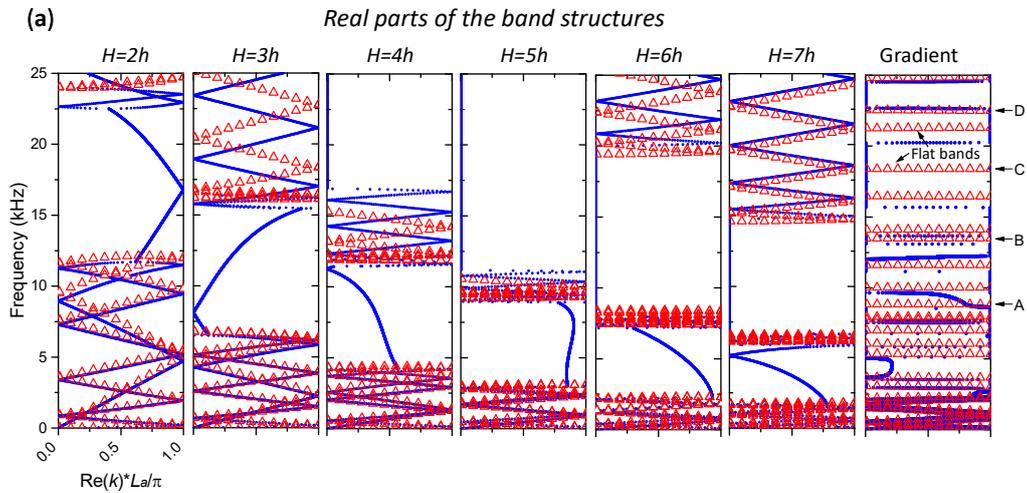

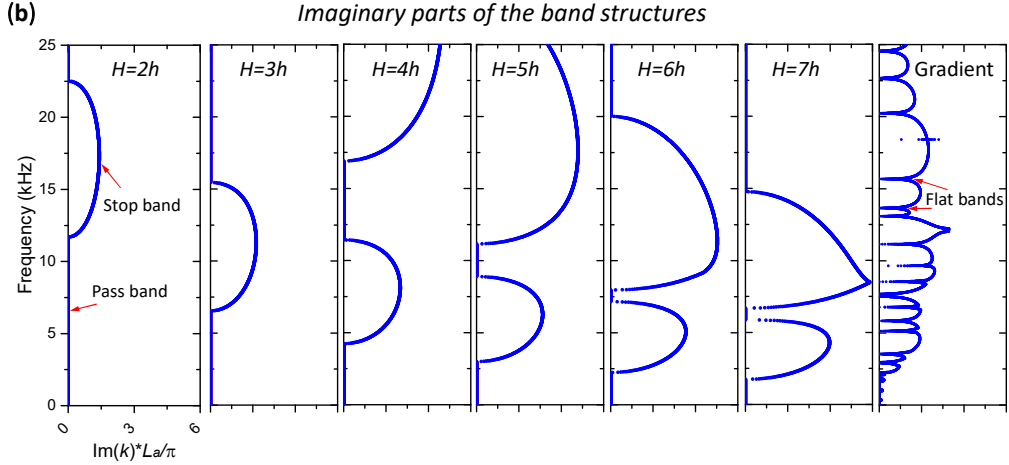

**Fig. 4**. Calculated (a) real parts of the band structures by the FEM (red triangles) and the SEM (blue dots) for the uniform and the gradient ($\alpha = 0.5$) cases, and (b) the corresponding imaginary parts by the SEM. In this figure, $L_a = 6a$.

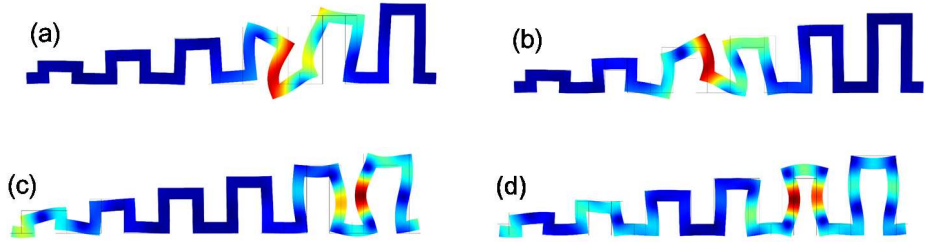

**Fig. 5**. Calculated eigen-modes by the FEM at different frequencies denoted by (a) arrow A ($f = 8.860$ kHz), (b) arrow B ($f = 11.629$ kHz), (c) arrow C ($f = 18.401$ kHz), and (d) arrow D ($f = 22.501$ kHz) in the rightmost sub-figure of **Fig. 4**(a). The color represents the amplitude of the total displacement.

*3.2 Effects of the gradient level $\alpha$ on the band structures*

**Fig. 6** shows the calculated imaginary parts of the band structures by the SEM for the gradient cases with different values of $\alpha$. The results indicate that the normal pass bands shrink with the increase of the $\alpha$ value, and are finally occupied by the flat bands. The results presented in **Fig. 6** clearly demonstrate that the gradient level $\alpha$ is the key factor determining the transformation of the band structures from the normal case to this special one in which the dispersion curves over a certain frequency are totally comprised of flat bands. Moreover, at the frequencies corresponding to the flat bands, certain parts of the metamaterial beam undergo a local resonance, which means that the wave energy will be localized in these parts.

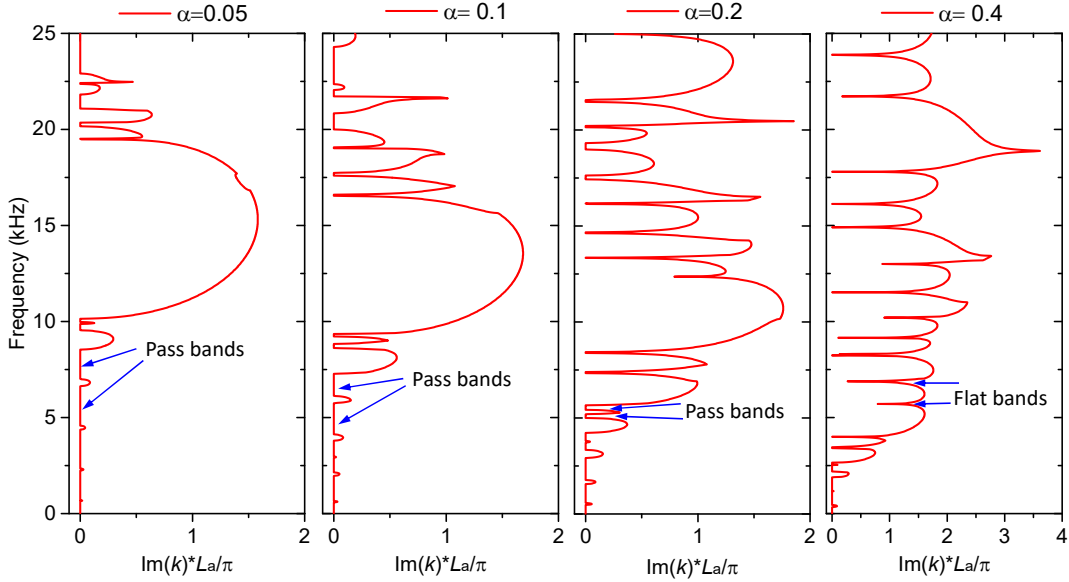

**Fig. 6**. Calculated imaginary parts of the band structures by the SEM for the gradient zigzag metamaterial beams with different gradient levels.

## 4  Gradient zigzag metamaterial beams as broadband vibration isolators

As mentioned above, the elastic waves with the frequencies at the flat bands have a zero group velocity, indicating that the elastic waves will be trapped by certain parts in the gradient zigzag metamaterial beams and cannot propagate. Consequently, based on the calculated band structures of the gradient cases in **Fig. 4** and **Fig. 6**, we can utilize our proposed gradient zigzag metamaterial beams to design broadband vibration isolators for beam-shaped structures.

*4.1 Numerical simulation of the wave transmission in gradient zigzag metamaterial beams*

In order to verify the aforementioned fact, the wave transmission spectrum of a finite gradient zigzag metamaterial beam ($\alpha = 0.5$) as shown in **Fig. 1**(a) is calculated by the FEM software COMSOL Multiphysics. The numerical model is established in **Fig. 7** in which the perfectly matched layers (PMLs) are applied to the left and right ends to avoid wave reflections, and a uniform displacement is prescribed on the line along the vertical direction to generate the incident flexural waves. The out-of-plane (vertical) displacement of the transmitted waves is then extracted to calculate the wave transmittance.

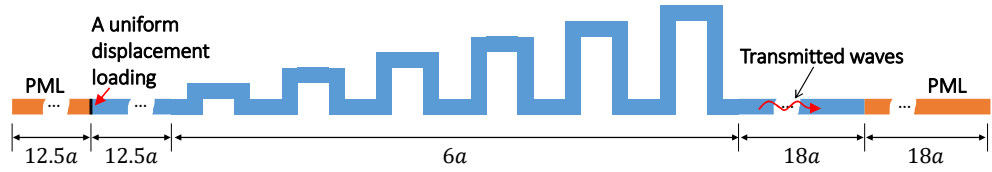

**Fig. 7**. The 2D numerical model to evaluate the wave transmission behavior of the gradient zigzag metamaterial beams by the FEM.

The wave transmittance is measured through the comparison of the amplitude of the vertical displacement of the transmitted waves with that of a straight beam of the thickness $h$. For comparison, the corresponding results for the uniform case are also calculated and presented in **Fig. 8**. As expected, the gradient zigzag metamaterial beam holds a low wave transmittance within a broad frequency range, i.e. [2.0, 25.0] kHz here, while the uniform case only works well as an isolator within the stop-bands (represented by the gray zones). It should be mentioned that although there exists a series of peaks in the transmission curve of the gradient case at the frequencies corresponding to the flat bands, their values are still much smaller than 1.0. This comparison confirms that the gradient zigzag metamaterial beam really works well as a broadband wave or vibration isolator within a broad frequency range of interest.

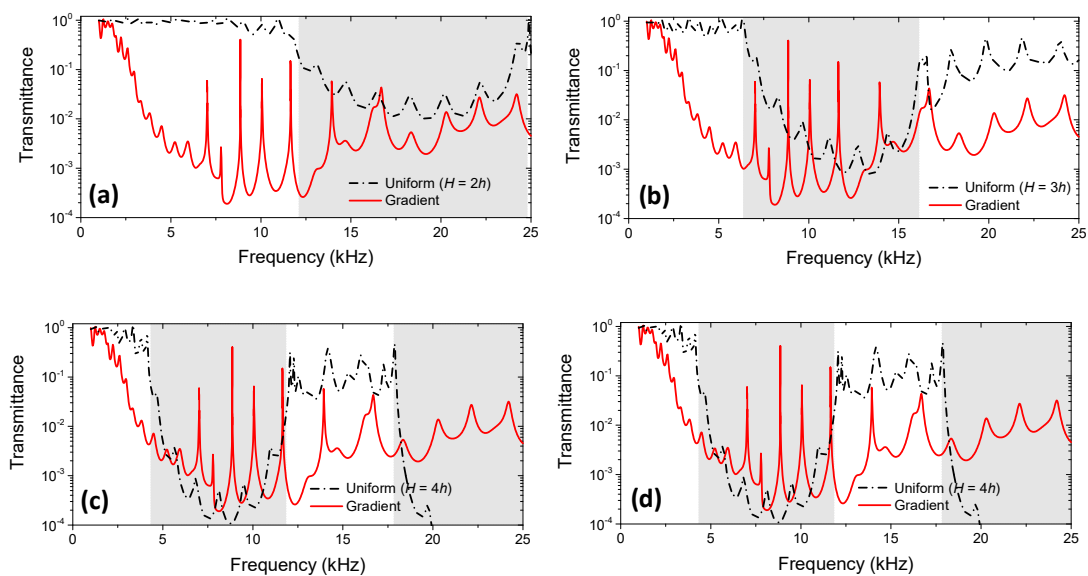

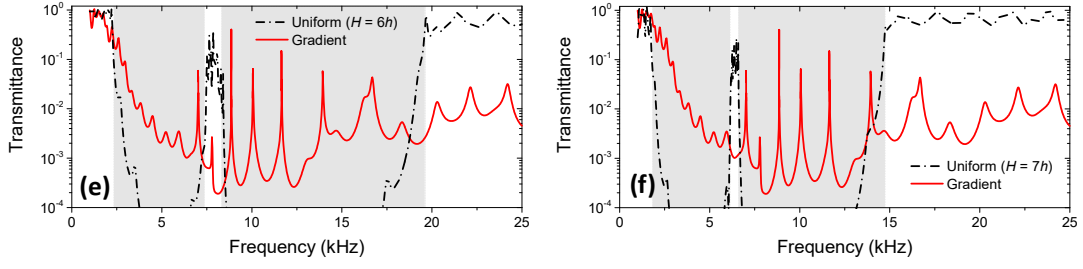

**Fig. 8**. Comparison of the numerically calculated transmittances for the gradient case (α = 0.5) and the uniform case with *H* equal (a) 2*h*, (b) 3*h*, (c) 4*h*, (d) 5*h*, (e) 6*h*, and (f) 7*h*. In this figure, the gray zones denote the stop-bands of the uniform cases.

*4.2 Experimental validation*

In order to validate the above numerical simulation results, the corresponding experiments were conducted based on the 3D-printed (J750, Stratasys, US) samples using the material VeroPureWhite. **Fig. 9**(a) shows the 3D-printed samples and their geometrical sizes and **Fig. 9**(b) shows the overall experimental setup. In the experiments, the incident waves are excited by two piezoelectric patches (Haiying Group, Wuxi, China) bonded on both the top and bottom surfaces of the beams, which are simultaneously excited by two opposite voltage signals. A damping material (blu-tack) is glued on the two ends of the samples to eliminate wave reflections from the boundaries. To measure the wave transmittance, a sweep signal of a duration of 0.04 s within [0.8, 26.0] kHz is generated by the build-in signal generator of the Laser Scanning Vibrometer (Polytec, PSV-500) and amplified by a power amplifier (ATA-308, Agitek, Xi'an, China), which is then input to the homemade phase conversion device. From the phase conversion device, two signals with opposite voltages are generated and input to the piezoelectric patches bonded on the top and bottom surfaces of the beams, respectively. The final voltages loaded on the piezoelectric patches are ±40 V. The transient out-of-plane velocities of the transmitted waves are measured by the Laser Scanning Vibrometer at the point 30 mm away from the top edge of the zigzag metamaterial beams. Then, the transient signals are transformed into the frequency domain by using the Fourier-transform. The measured results for the straight beam of the thickness *h* are taken as the reference results. The wave transmittances for the uniform and gradient metamaterial beams are then calculated through the comparison of the magnitudes of the measured out-of-plane velocities with those of the reference.

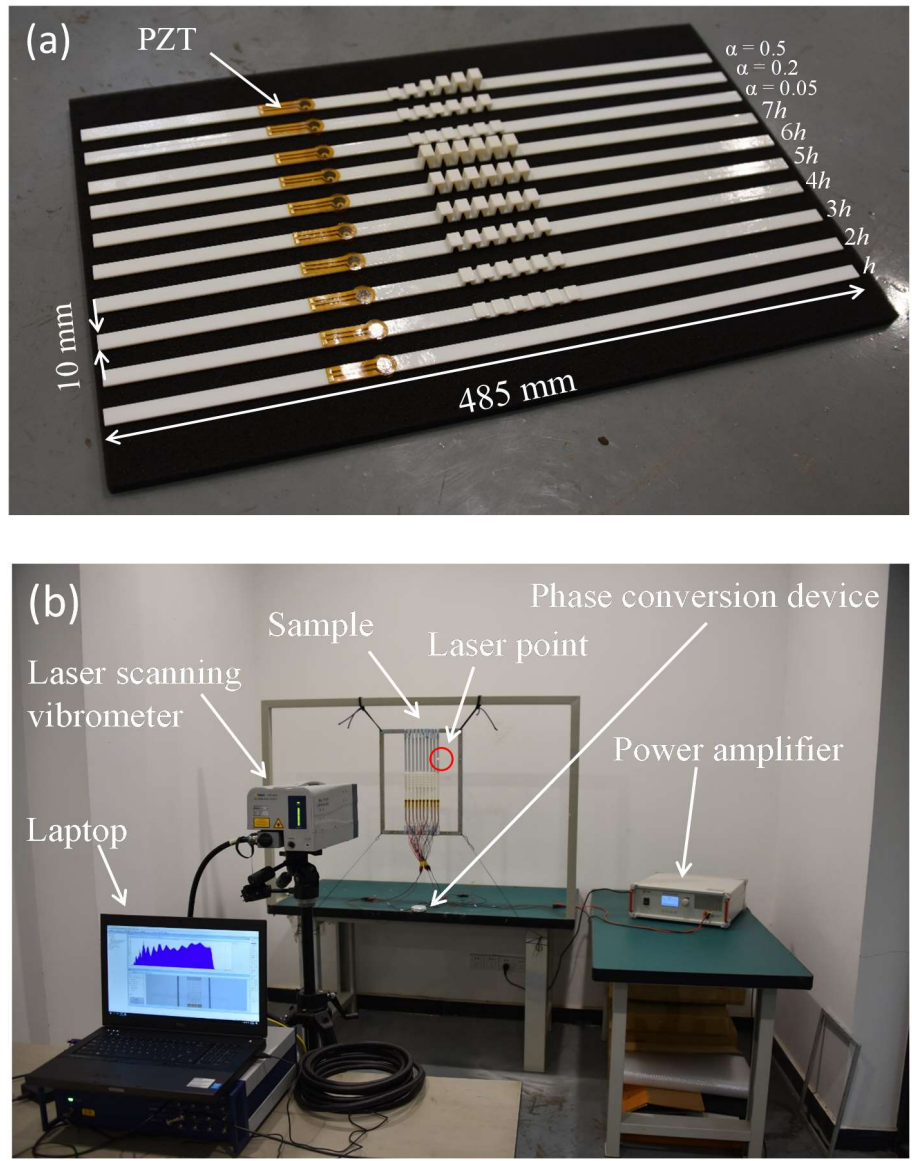

**Fig. 9**. Experimental validation of the numerical simulations in the Subsection 4.1. (a) The 3D-printed experimental samples. (b) The overall experimental setup.

The comparison of the transmittances obtained by the experiments and the numerical simulations is presented in **Fig. 10**. The comparison shows that the experimental results agree fairly well with the numerical simulation results, and the general trends of the numerical simulations are well captured by the experiments. In addition, through the comparison of the results in **Fig. 10**(g)-(i), it is found that the frequency range in which the transmitted wave has a low transmittance becomes wider with the increase of the α value. This confirms the key role of the gradient level α in the proposed gradient zigzag metamaterial beams as broadband wave or vibration isolators again. In general, the results presented in **Fig. 10** firmly confirm the good performance of our proposed gradient metamaterial beams as potential broadband vibration isolators for beam-shaped structures.

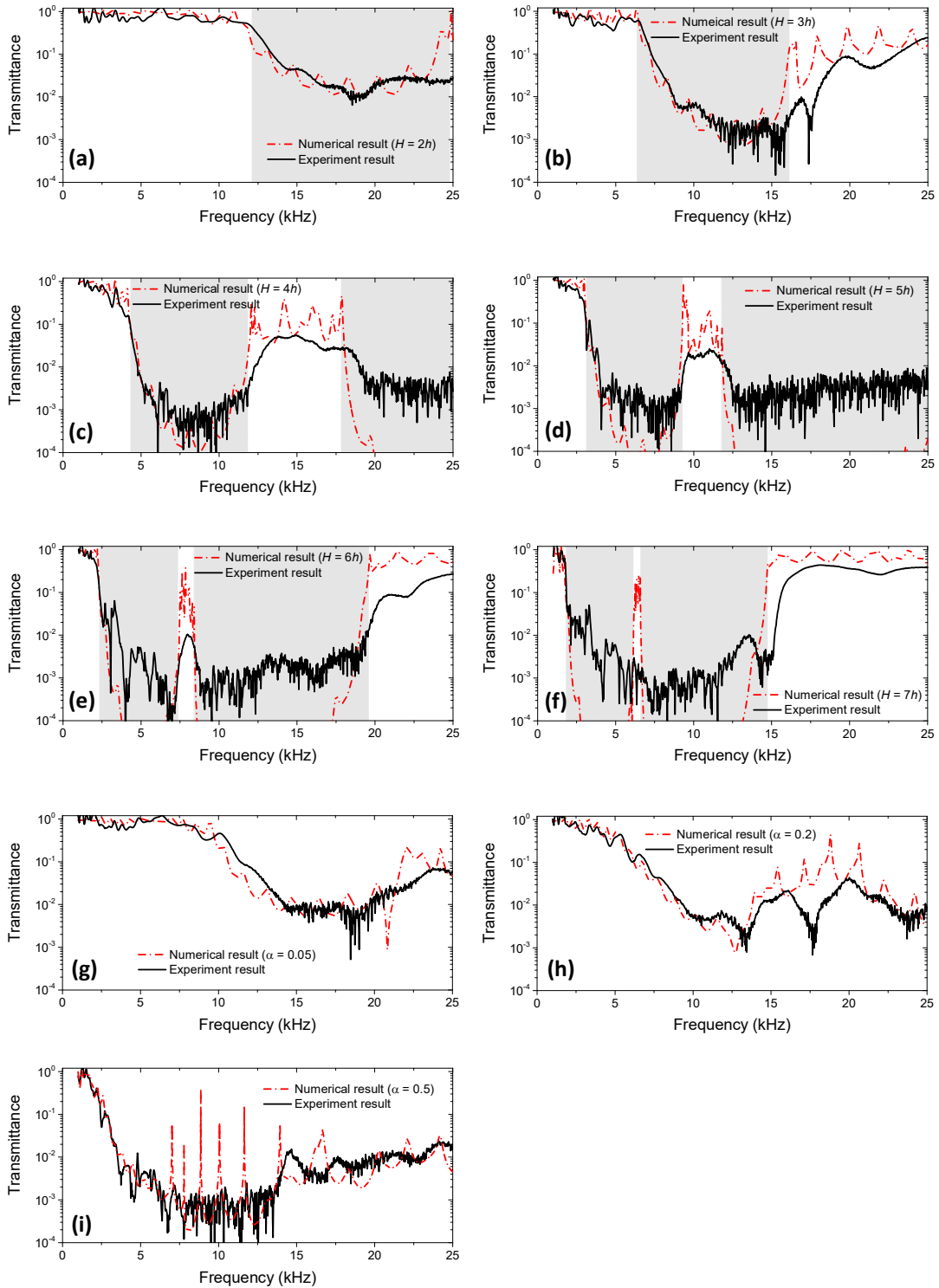

**Fig. 10**. Comparison of the transmittances obtained by the experiments and the numerical simulations for the uniform cases with (a) *H*=2*h*, (b) *H*=3*h*, (c) *H*=4*h*, (d) *H*=5*h*, (e) *H*=6*h*, (f) *H*=7*h*, and the gradient cases with (g) α = 0.05, (h) α = 0.2, and (i) α = 0.5, respectively. In this figure, the gray zones denote the stop-bands of the uniform cases the same as those in **Fig. 8**.

## 5. Conclusions

In this work, we propose a novel kind of gradient zigzag metamaterial beams as our previously proposed uniform counterparts. The calculated real and complex band structures by the FEM and the SEM indicate that such gradient metamaterial beams have an essentially different property from the uniform case, namely, their band structures are exclusively composed of flat bands within a broad frequency range. These flat bands are caused by the local resonance of certain parts in the gradient zigzag metamaterial beams, and the elastic waves at these flat bands will be trapped by the resonant parts and cannot propagate. As an illustrative and straightforward application, our proposed gradient zigzag metamaterial beams can be utilized as broadband wave or vibration isolators for beam-like structures. As a proof, the transmission spectra of our proposed gradient zigzag metamaterial beams are numerically calculated and verified by experiments based on the 3D-printed samples. The experimental results are in fairly good agreement with the numerical simulation results, and both results show that our proposed gradient zigzag metamaterial beams have in fact a low wave transmittance within a broad frequency range of interest. This work paves a new way to design broadband wave or vibration isolators.

**Acknowledgments**

This work was supported by the National Natural Science Foundation of China (Nos. 12072051, 11632004 and U1864208), the Venture and Innovation Support Program for Chongqing Overseas Returnees (cx2018050), the Fundamental Research Funds for the Central Universities (No.300102251513), the National Science and Technology Major Project (2017-VII-0011-0106), and the Science and Technology

Planning Project of Tianjin (20ZYJDJC00030). The funding provided by the China Scholarship Council (CSC) to support Jun Zhang (201806055009) as a Visiting Scholar at the Chair of Structural Mechanics, University of Siegen, Germany, is also gratefully acknowledged.

waves in thin plates by 3D-printed metasurfaces, Journal of Sound and Vibration, under review (2019).

[38] U. Lee, Dynamics of Beams and Plates, in: Spectral Element Method in Structural Dynamics, 2009, pp. 111-131.

[39] P. Wang, T.N. Chen, K.P. Yu, X.P. Wang, Lamb wave band gaps in a double-sided phononic plate, Journal of Applied Physics, 113 (2013).

[40] G. Wang, D.L. Yu, J.H. Wen, Y.Z. Liu, X.S. Wen, One-dimensional phononic crystals with locally resonant structures, Physics Letters A, 327 (2004) 512-521.

[41] Z.W. Zhu, Z.C. Deng, Identical band gaps in structurally re-entrant honeycombs, Journal of the Acoustical Society of America, 140 (2016) 898-907.

**Supplementary material 1**: Measurement of Young's modulus of the 3D-printing material VeroPureWhite

The wavelength of flexural waves in a 485×200×2 mm plate printed using VeroPureWhite has been measured at several different frequencies and is shown in **Figure S1**. Then, the measured data are fitted using the theoretical formula $\lambda_F = 2\pi * \left[\frac{Eh^2}{12\,(1-\nu^2)\omega^2}\right]^{1/4}$ for the wavelength of flexural waves in a plate with the fitting parameter $E$, which represents Young's modulus of the plate. Other material parameters of VeroPureWhite are the mass density $\rho = 1185$ kg/m$^3$ and Poisson's ratio $\nu = 0.4185$, which were measured in our previous paper [37]. The fitted Young's modulus of VeroPureWhite is $E = 3.2394$ GPa.

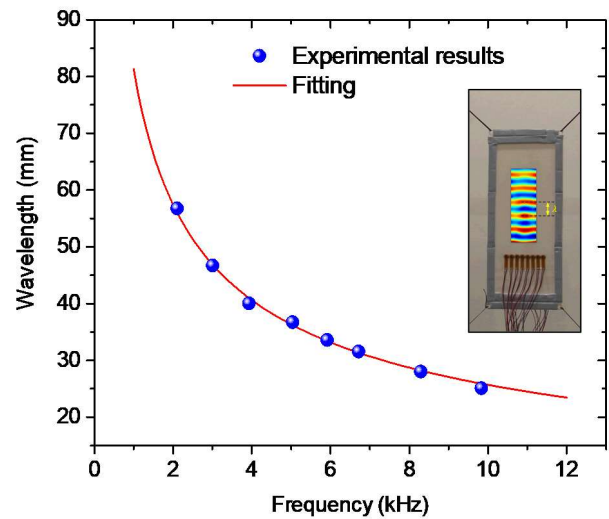

**Figure S1**. Measured wavelength (blue solid dots) of flexural waves at several different frequencies based on the scanned full velocity fields using Polytec PSV-500 and its numerical fitting curve (solid red line). The wavelength has been evaluated through the measurement of the distance between two adjacent crests or troughs of the measured velocity fields.

**Supplementary material 2**: The detailed expressions for the coefficient matrix in Eq. (6)

$$\mathbf{D}_2 = -S_{12}S_{22}^{-1}S_{23} + S_{13}$$
$$\mathbf{D}_1 = -S_{12}S_{22}^{-1}S_{21} + S_{11} - S_{32}S_{22}^{-1}S_{23} + S_{33}$$
$$\mathbf{D}_0 = -S_{32}S_{22}^{-1}S_{21} + S_{31}$$